\documentclass[runningheads]{llncs}
\usepackage[T1]{fontenc}
\usepackage{graphicx}
\usepackage{amsmath}
\begin{document}
\title{Hallucination Index: An Image Quality Metric for Generative Reconstruction Models}

\titlerunning{Hallucination Index}
%
\author{Matthew Tivnan\inst{1} 
\and
Siyeop Yoon\inst{1} 
\and
Zhennong Chen\inst{1} 
\and
Xiang Li\inst{1} 
\and
Dufan Wu\inst{1} 
\and
Quanzheng Li\inst{1} 
}
\authorrunning{M. Tivnan et. al.}
%
\institute{Massachusetts General Hospital and Harvard Medical School, 55 Fruit Street, Boston, MA, 02114. \email{li.quanzheng@mgh.harvard.edu}}

\maketitle              
%

\begin{abstract}
Generative image reconstruction algorithms such as measurement conditioned diffusion models are increasingly popular in the field of medical imaging. These powerful models can transform low signal-to-noise ratio (SNR) inputs into outputs with the appearance of high SNR. However, the outputs can have a new type of error called hallucinations. In medical imaging, these hallucinations may not be obvious to a Radiologist but could cause diagnostic errors. Generally, hallucination refers to error in estimation of object structure caused by a machine learning model, but there is no widely accepted method to evaluate hallucination magnitude. In this work, we propose a new image quality metric called the hallucination index. Our approach is to compute the Hellinger distance from the distribution of reconstructed images to a zero hallucination reference distribution. To evaluate our approach, we conducted a numerical experiment with electron microscopy images, simulated noisy measurements, and applied diffusion based reconstructions. We sampled the measurements and the generative reconstructions repeatedly to compute the sample mean and covariance. For the zero hallucination reference, we used the forward diffusion process applied to ground truth. Our results show that higher measurement SNR leads to lower hallucination index for the same apparent image quality. We also evaluated the impact of early stopping in the reverse diffusion process and found that more modest denoising strengths can reduce hallucination. We believe this metric could be useful for evaluation of generative image reconstructions or as a warning label to inform radiologists about the degree of hallucinations in medical images. 
\keywords{Hallucinations, Uncertainty, Diffusion Models, Generative Models, Medical Image Reconstruction}
\end{abstract}
\section{Introduction}

In the rapidly evolving landscape of medical image reconstruction algorithms, generative models have become increasingly popular. Examples  include generative adversarial networks \cite{suganthi2021review,yi2019generative,zhou2023gan}, normalizing flow models \cite{denker2020conditional} 
\cite{hajij2022normalizing}, and diffusion models  \cite{kazerouni2023diffusion} \cite{khader2023denoising}\cite{song2021solving}\cite{xie2022measurement}. These generative models have shown exceptional capability in transforming low signal-to-noise ratio (SNR) inputs into higher quality outputs that appear to have high SNR with realistic structural details. However, the advancement of generative reconstruction methods introduces a new type of error called hallucinations. Hallucinations are commonly referenced in research literature \cite{barbano2023steerable} \cite{bhadra2021hallucinations} \cite{yi2019generative} \cite{mardani2018deep} \cite{cohen2018distribution} but definitions vary. Generally, hallucinations are structural uncertainty in the output of machine learning models, especially generative models. It is distinct from measurement noise which does not have the appearance of varied object structure.

Hallucinations in the context of medical imaging are particularly concerning and several members of the field are sounding the alarm \cite{chu2020potential} \cite{teneggi2023trust}. These errors, while not immediately obvious to radiologists, could lead to diagnostic inaccuracies, compromising patient safety. Despite the recognition of hallucinations as a potential issue, the field lacks a standard methodology for quantitative evaluation of such generative artifacts. To address this gap, we propose the \textit{hallucination index}, a novel image quality metric designed to evaluate the extent of hallucinations in the outputs of generative  algorithms. Our approach employs the Hellinger distance \cite{nikulin2001hellinger} between the distribution of reconstructed images and a zero-hallucination reference distribution. The reference distribution should be generated without a machine learning model, and it should have the same apparent SNR as the reconstructions without the structural variability. For example, in diffusion models, we can apply the forward diffusion process to ground truth.

One noteworthy alternative for evaluating hallucinations in medical images is the \textit{Hallucination Map} proposed by \cite{bhadra2021hallucinations}. This method involves projecting the reconstructions to the imaging system null space to identify features which provably did not arise from the measurements. While this method could be extremely useful, especially for ill-conditioned inverse problems, we argue it does not capture the possibility of hallucinations within the measurement space. For example, if an imaging system is unbiased with additive noise, it technically has no null space, but a generative denoising model may still be affected by hallucinations.  By comparison, our hallucination index is a distribution-to-distribution distance metric, which could measure the hallucinations in this scenario. 


The  image reconstruction task we simulated is electron microscopy (EM) of cortical neurons using the Machine Intelligence from Cortical Networks (MICrONS) dataset \cite{microns2021functional} which is openly available under a Creative Commons Public License. This modality offers the high spatial resolution necessary for neuron body segmentation and connectome mapping. However, the time it takes to scan hundreds of terabytes of imaging data per cubic millimeter represents a significant bottleneck. One proposed solution is to decrease the dwell time \cite{buban2010high}. While effective in reducing scan time, this method decreases the SNR of the resulting images \cite{trampert2018should}. Generative image restoration with diffusion models emerges as a promising solution for low dwell time EM reconstruction, but one concern is that hallucinated artifacts could interfere with segmentation, leading to incorrect interpretations of neural structures. Our experiment aims to evaluate the tendency of diffusion based reconstructions to hallucinate and how those hallucinations vary with respect to other parameters and metrics like input SNR, reverse process stopping time, and mean squared error.

\section{Theoretical Methods}
\begin{figure}
    \centering
    \includegraphics[width=1.0\textwidth]{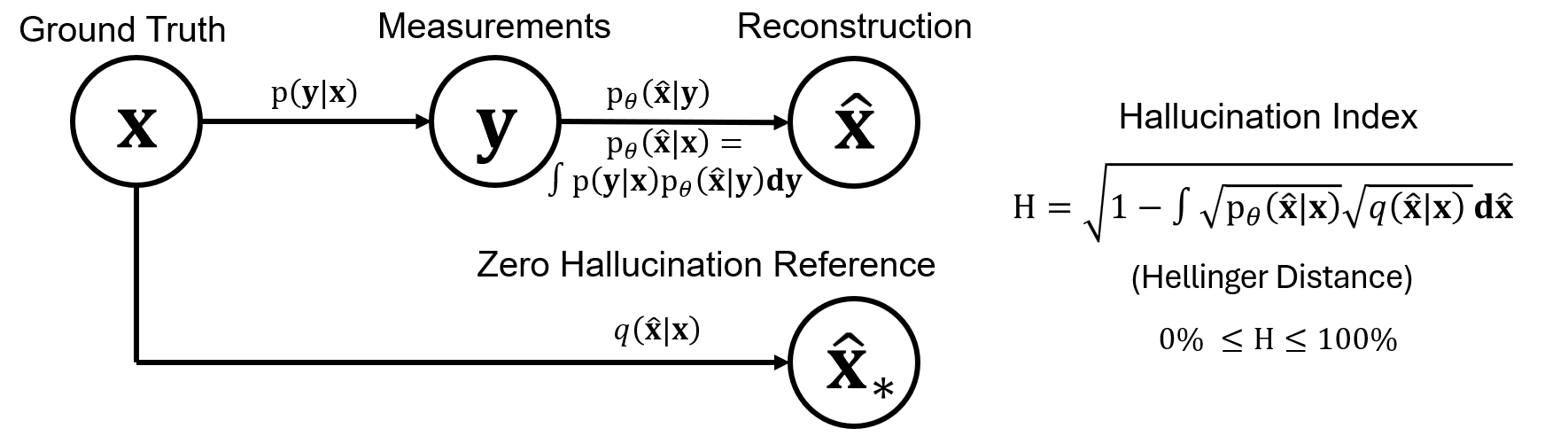}
    \caption{Probabilistic graphical model of true images, measurements, and reconstructions. The hallucination index is defined as the Hellinger distance between the reconstruction distribution and a zero-hallucination reference distribution. In diffusion models, the forward process can be used as the zero hallucination reference.}
    \label{fig:hallucination_index_pgm}
\end{figure}

\subsection{Hallucination Index}

A probabilistic model of image reconstruction is shown in Figure 1. We consider the true images, $\mathbf{x}$, the measurements, $\mathbf{y}$, and the reconstructed images, $\mathbf{\hat{x}}$, as flattened random vectors. The measurements are a stochastic function of the true images, related by the conditional probability density function $p(\mathbf{y}|\mathbf{x})$. Classical reconstruction methods are often deterministic functions of the measurements, including linear deterministic methods like pseudo-inverse reconstruction or nonlinear deterministic methods like pre-trained convolutional neural networks. However, there is increasing interest in generative reconstruction methods, such as diffusion models, which are a stochastic function of the measurements. Therefore, we model the reconstruction process with the conditional probability distribution $p_{\boldsymbol{\theta}}(\mathbf{\hat{x}}|\mathbf{y})$, dependent on some parameter vector, ${\boldsymbol{\theta}}$. The joint probability density function is

\begin{equation}
p(\mathbf{x},\mathbf{y}, \mathbf{\hat{x}}) = p(\mathbf{x})p(\mathbf{y}|\mathbf{x}) p_{\boldsymbol{\theta}}(\mathbf{\hat{x}}|\mathbf{y}) .
\label{eq:joint_pdf}
\end{equation}



Therefore, an end-to-end conditional probability density function of reconstructions given true images is given by

\begin{equation}
p_{\boldsymbol{\theta}}(\mathbf{\hat{x}}|\mathbf{x}) = \int  p(\mathbf{y}|\mathbf{x}) p_{\boldsymbol{\theta}}(\mathbf{\hat{x}}|\mathbf{y}) \mathbf{dy}
\end{equation}

This distribution of reconstructions can contain various types of error such as bias (e.g. spatial blur), noise, and hallucinations.  In the context of image reconstruction, hallucinations are generally understood to refer to variability or uncertainty in the structure of an object arising from the application of a machine learning based image processing model. For example, if the measurement signal-to-noise ratio (SNR) is low, and a generative image reconstruction method is repeatedly applied to produce multiple samples of output images with the appearance of high SNR, this will typically result in highly varied estimates of the object underlying the noise. 

The key to our approach is to define what hallucinations are not. We claim that when no machine learning model has been applied, there are no hallucinations. For example, if white noise is added to the input, there will be increased variance in the output, but that error is not a hallucination by definition because no machine learning model has been applied. Therefore, we introduce a new distribution, $q(\mathbf{\hat{x}}|\mathbf{x})$ which we define as a zero hallucination reference with the same apparent signal-to-noise ratio as the generative reconstructions $p_{\boldsymbol{\theta}}(\mathbf{\hat{x}}|\mathbf{x})$. In the case of diffusion models, this zero hallucination reference is available via the forward diffusion process. We define the \textit{hallucination index} as the Hellinger distance between the distribution of reconstructed images and the zero hallucination reference distribution as shown

\begin{equation}
H\big(p_{\boldsymbol{\theta}}(\mathbf{\hat{x}}|\mathbf{x}), q(\mathbf{\hat{x}}|\mathbf{x})\big) = \sqrt{1 - \int \sqrt{p_{\boldsymbol{\theta}}(\mathbf{\hat{x}}|\mathbf{x})} \sqrt{ q(\mathbf{\hat{x}}|\mathbf{x}))} \mathbf{d\hat{x}}}  
\label{eq:hallucination_index}
\end{equation}

This metric has several desirable mathematical properties. First of all, it is bounded by $0\leq H \leq 1$ with equality to zero when the two distributions are identical and equality to one when the two distributions have no overlap. Loosely, the hallucination index can be interpreted as the percentage of error due to hallucinations as opposed to noise. The Hellinger distance is a metric that satisfies the triangle inequality which could be useful for analyzing a cascade of multiple data processing steps.  Rearranging terms in \eqref{eq:hallucination_index} results in

\begin{equation}
H\big(p_{\boldsymbol{\theta}}(\mathbf{\hat{x}}|\mathbf{x}), q(\mathbf{\hat{x}}|\mathbf{x})\big) = \sqrt{1 - \int p_{\boldsymbol{\theta}}(\mathbf{\hat{x}}|\mathbf{x}) \frac{\sqrt{ q(\mathbf{\hat{x}}|\mathbf{x}))}}{\sqrt{p_{\boldsymbol{\theta}}(\mathbf{\hat{x}}|\mathbf{x})} } \mathbf{d\hat{x}}}  
\label{eq:hallucination_index_2}
\end{equation}

\noindent The form of \eqref{eq:hallucination_index_2}  now includes an expectation over $p_{\boldsymbol{\theta}}(\mathbf{\hat{x}}|\mathbf{x})$. This mean can be approximated by the sample mean, and the probability densities can be approximated by various methods such as kernel density estimation. In this work, we chose to use a simpler approach where the hallucination index is approximated by the Hellinger distance between two multivariate normal distributions with the same mean and covariance as $p_{\boldsymbol{\theta}}(\mathbf{\hat{x}}|\mathbf{x})$ and $q(\mathbf{\hat{x}}|\mathbf{x})$ using the closed form solution

\begin{gather}
H\approx \sqrt{1 - \exp{\Big(-\frac{1}{4}(\boldsymbol{\mu}_p - \boldsymbol{\mu}_q)^T ({\boldsymbol{\Sigma}_p + \boldsymbol{\Sigma}_q})^{-1}(\boldsymbol{\mu}_p - \boldsymbol{\mu}_q) + c \Big)}}\label{eq:hallucination_index_gaussian}
\\
c = \frac{1}{4} \log |\boldsymbol{\Sigma}_p| + \frac{1}{4} \log |\boldsymbol{\Sigma}_q| - \frac{1}{4} \log |{\boldsymbol{\Sigma}_p + \boldsymbol{\Sigma}_q}| \nonumber
\end{gather}

We approximate the covariance matrices by a circulant covariance matrix with the same noise power spectrum. So, in practice we compute the hallucination index by 1) taking multiple samples of the measurements 2) applying the generative reconstruction model 3) computing the sample mean of the reconstructions 4) compute sample noise power spectrum of the reconstruction distribution and the zero hallucination reference distribution 5) compute the hallucination index using the formula above.

\subsection{Fourier Diffusion Models}

We followed the methods in \cite{tivnan2023fourier} to implement Fourier diffusion models. The main difference with respect to most diffusion models is that the scalar multiplication and additive white Gaussian noise in the forward stochastic process is generalized to include convolutional operators and additive stationary Gaussian noise. The practical benefit of this method is that we can construct a diffusion bridge model starting from ground truth images at $t=0$ and converging to the same noise power spectrum as the measurements. At inference time, the reverse process can be initialized at $t=t_\text{start}$ directly from the measurements. Fourier diffusion models are based on a forward stochastic process defined by

\begin{equation}
\mathbf{x}_t = \mathbf{H}_t \mathbf{x}_0 + \boldsymbol{\Sigma}_t^{1/2} \frac{1}{\sqrt{t}} \mathbf{w}_t, \quad t>0
\end{equation}

\noindent where $\mathbf{H}_t$ is a invertible circulant matrix operator representing the modulation transfer function in the spatial frequency domain, $\boldsymbol{\Sigma}_t$ is a circulant covariance matrix of spatially stationary Gaussian noise, and $\mathbf{w}_t$ is a standard Wiener process. The corresponding forward diffusion stochastic differential equation is

\begin{gather}
   \mathbf{dx}_t = \mathbf{F}_t \mathbf{x}_t dt + \mathbf{G}_t \mathbf{dw}_t, \enspace \mathbf{F}_t = \mathbf{H}^{'}_t \mathbf{H}_t^{-1}, \enspace 
   \mathbf{G}_t = [\boldsymbol{\Sigma}_t^{'} - \mathbf{F}_t \boldsymbol{\Sigma}_t  -  \boldsymbol{\Sigma}_t \mathbf{F}_t^T  ]^{1/2}
\end{gather}

Following the framework for more general stochastic differential equations in Appendix A of \cite{song2020score}, we can define the reverse stochastic differential equation as

\begin{gather}
   \mathbf{dx}_t = [\mathbf{F}_t \mathbf{x}_t - \frac{1}{2} \mathbf{G}_t \mathbf{G}_t^T \nabla_{\mathbf{x}_t}\log p(\mathbf{x}_t)] dt + \mathbf{G}_t \mathbf{d\bar{w}}_t . \label{eq:reverse_process}
\end{gather}

The generative model is implemented by approximating the score function with a score matching neural network, $\mathbf{s}_{\theta}(\mathbf{x}_t, t) \approx \nabla_{\mathbf{x}_t}\log p(\mathbf{x}_t)$. We can sample from the reverse process, $p_{\theta}(\mathbf{x}_t|\mathbf{x}_{t_\text{start}} = \mathbf{y})$  using stochastic integral sampling methods such as the Euler-Maruyama method applied to \eqref{eq:reverse_process}. For hallucination analysis, this will be compared to the forward process, $q(\mathbf{x}_t|\mathbf{x}_0)\sim \mathcal{N}(\mathbf{H}_t\mathbf{x}_0,\boldsymbol{\Sigma}_t)$, which is a zero-hallucination reference with the same image quality in terms of modulation transfer function and noise power spectrum.




\begin{figure}[hb!]
    \centering
    \includegraphics[width=1.0\textwidth]{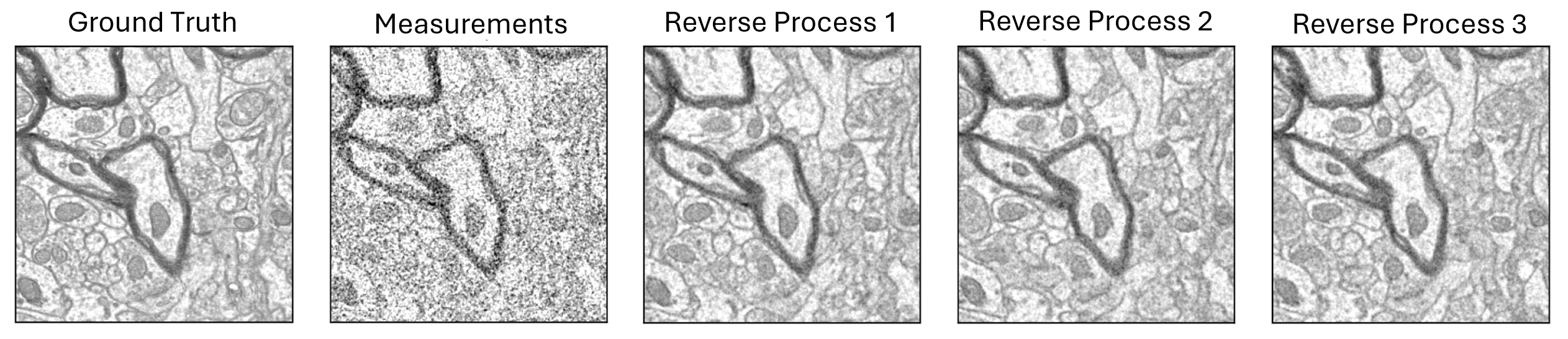}
    \caption{Ground truth, measurements, and reverse process samples with hallucinations.}
    \label{fig:enter-label}
\end{figure}

\section{Experimental Methods}

We conducted a large scale numerical experiment to evaluate hallucinations in Fourier diffusion models with various configurations. The purpose of this experiment is twofold. First, we seek to validate our proposed metric to verify that it corresponds to what is commonly understood as hallucinations by visual inspection. Second, we seek to quantify the relationship between hallucination index and other measurable parameters and metrics such as input SNR, reverse process early stop time, noise power spectrum, and mean squared error. 

The experiment simulates the task of imaging cortical neurons through electron microscopy, particularly under conditions where the images are affected by low SNR due to faster scanning. We used the  Machine Intelligence from Cortical Networks (MICrONS) dataset to model the ground truth images \cite{microns2021functional}.  We used 10,000 image samples of image size 256x256 for training, 1,000 samples for validation, and 1,000 samples for evaluation. 

\begin{figure}[ht!]
    \centering
    \includegraphics[trim={49mm 42mm 53mm 42mm},clip,width=\textwidth]{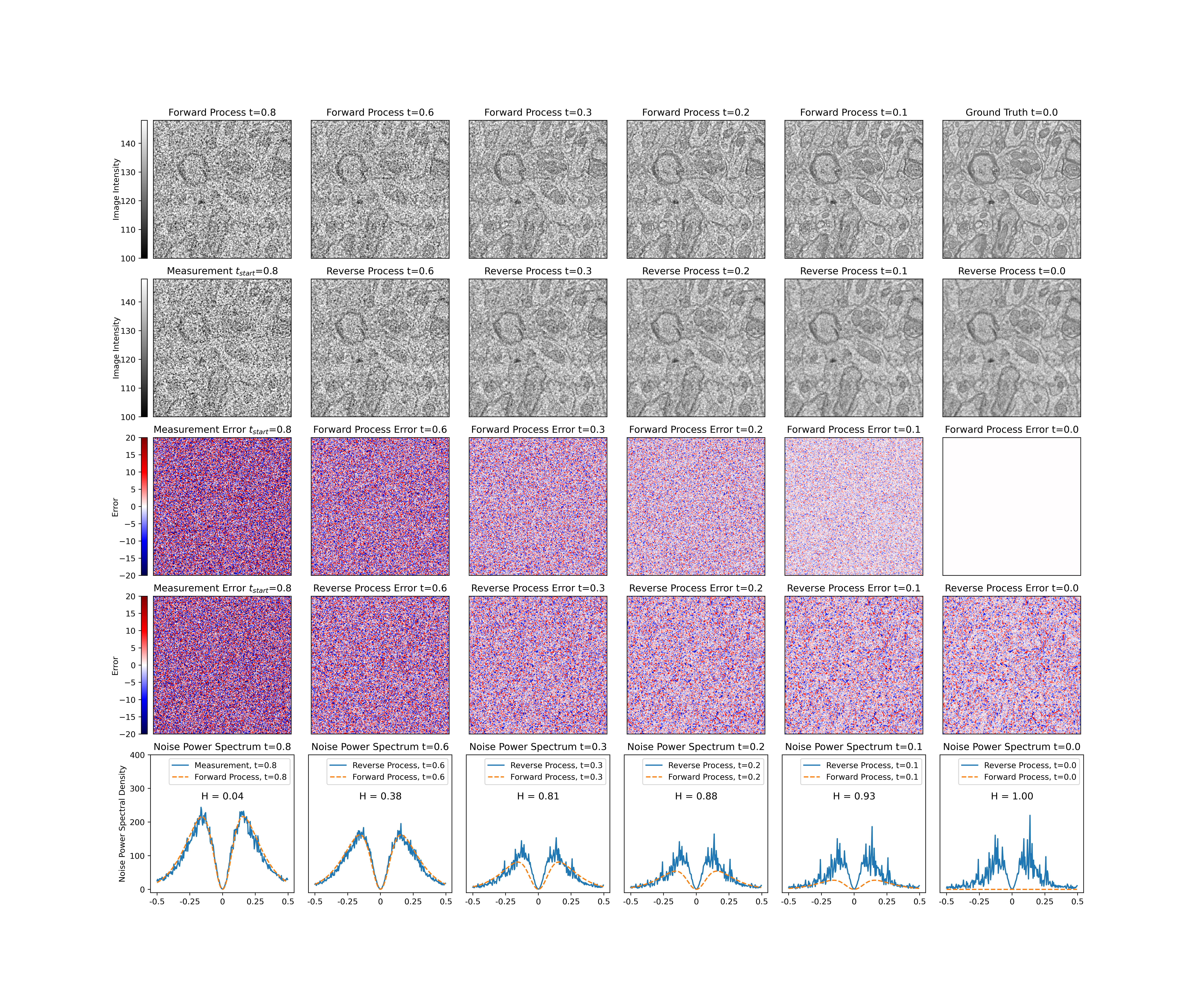}
    \caption{ Comparison between forward and reverse process samples.  Hallucination index increases for more reverse diffusion steps because there is increased Hellinger distance between forward and reverse process.    }
    \label{fig:forward_reverse_NPS}
\end{figure}

For each evaluation case, we simulated 64 measurement noise realizations. The measurement model is the ground truth images plus additive stationary Gaussian noise with noise power spectrum, $\boldsymbol{\Sigma}_t$, as shown in Figure \ref{fig:forward_reverse_NPS}. For the Fourier diffusion model, we did not include any spatial blur, so $\mathbf{H}_t=\mathbf{I}$, and the noise covariance increases linearly from zero noise to the same noise power spectrum as the measurements. This is similar to the variance-exploding stochastic process defined in \cite{song2020score} but with spatially correlated noise rather than white noise.

For training the Fourier diffusion model, we used the exact same methods described in \cite{tivnan2023fourier} including score matching neural network architecture and score matching loss function. We used 1,000 training epochs, 100 iterations per epoch, 32 images per batch, and the Adam optimizer with learning rate $10^{-3}$. After initializing with measurements, we sample the reverse process with the Euler-Maruyama method with 100 equally spaced time steps. We repeated the reverse diffusion process sampling for each evaluation sample. An example of multiple reverse process samples is shown in Figure \ref{fig:forward_reverse_NPS}.  We also repeated all experiments for 5 different measurement signal-to-noise ratio levels. This was accomplished by initializing the reverse process with measurement simulations corresponding to the intermediate noise power spectrum at $t_\text{start}$ values of 1.0, 0.8, 0.6, 0.4, and 0.2. When running the reverse diffusion sampling, we saved all the intermediate time step results. All together, this forms a large scale dataset that we used for image quality evaluation. Each reconstruction distribution in the dataset represents 64 samples from $p_{\theta}(\mathbf{\hat{x}}|\mathbf{x})$. We computed the mean, bias, variance, mean squared error, and noise power spectrum. We computed the hallucination index as a function of bias and noise power spectrum using \eqref{eq:hallucination_index_gaussian}.


\section{Results}

Figure \ref{fig:forward_reverse_NPS} shows a comparison of the forward and reverse process. The forward process does not have hallucinations, it is ground truth plus noise. The reverse process has hallucinations introduced by the diffusion model. The error images show increasing divergence between the reverse process and the forward process with more reverse diffusion steps. Visually, the error images for the reverse process error near $t=0$ show increasing intensity of structural patterns. Hallucination is increasing with more reverse diffusion steps because there is increased Hellinger distance between the forward and reverse process distributions.

   

\begin{figure}[ht!]
    \centering
    \includegraphics[width=\textwidth]{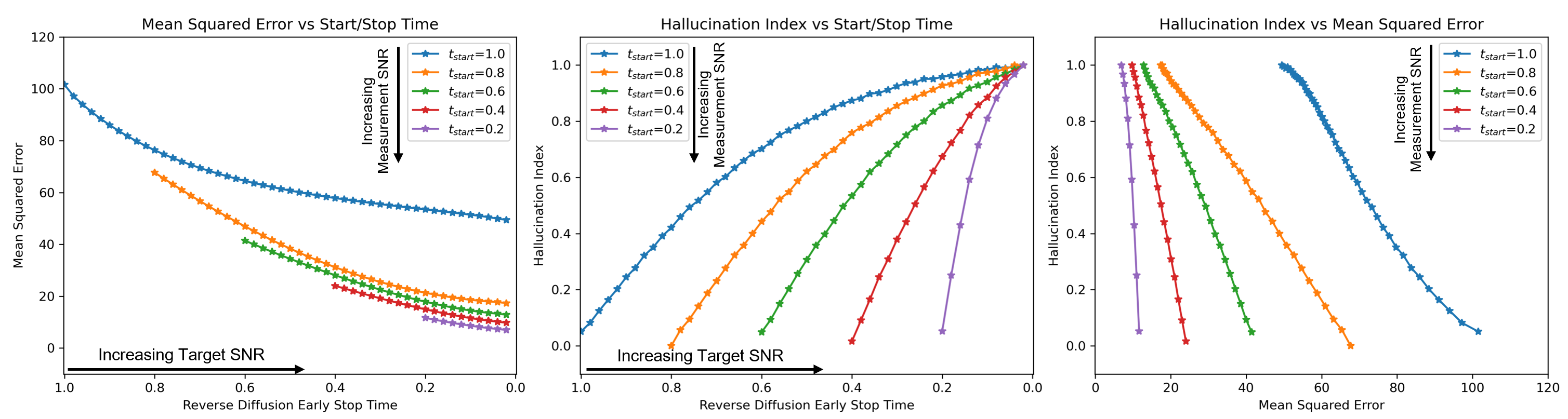}
    
       \caption{Image quality metrics as a function of input SNR and reverse process time step. Increasing measurement SNR decreases both mean squared error and hallucinations. Increasing reverse process time steps reduces mean squared error but increases hallucinations. Third plot shows tradeoff between mean squared error and hallucinations.}

   \label{fig:quantitative_metrics}
\end{figure}

Figure \ref{fig:quantitative_metrics} shows a quantitative analysis of mean squared error and hallucination index as a function of measurement SNR and reverse diffusion time steps. Both mean squared error and hallucination index are reduced if measurement SNR is increased. However, this is not always an option as it requires improvements to the imaging data acquisition. Mean squared error can also be reduced by running more iterations of a reverse diffusion process. However, we see that hallucination index increases with more reverse diffusion time steps. Overall there is a tradeoff between mean squared error and hallucination index. That is, the more aggressive the denoising applied by a generative reconstruction model, the more it hallucinates.


\section{Conclusion}

Our results demonstrate how the proposed hallucination index can be useful for identifying and quantifying generative artifacts that are not sufficiently described by mean squared error. However, our approach is not without its limitations. Notably, we assumed that the reconstruction distribution is Gaussian and shift-invariant. While these approximations facilitate computation, they may not fully capture the complexity of the error distributions. We also acknowledge that we have only tested our method on Fourier diffusion models, which form a direct diffusion bridge from measured images to noiseless images. One unique aspect of these models is the capacity for early stopping. This property was critical for our work because we used the forward process at the same intermediate time as the zero hallucination reference. This highlights a limitation of our approach that for generative models purporting to synthesize noiseless images given noisy inputs, the hallucination index will invariably evaluate to 100\%, due to the fact that noiseless images are given by a deterministic Dirac delta distribution. However, we note that 100\% hallucination index does accurately reflect the percent of covariance that is attributable to hallucinations as opposed to noise, for cases purporting to be a noiseless reconstruction. 

Our approach relies on a well-defined zero hallucination reference distribution, and we use the Hellinger distance to measure the distance to the distribution of reconstructions. In the future, we are interested to investigate other options for the zero hallucination reference distribution for cases when the forward diffusion process may not be available. We can also investigate alternative distance metrics such as KL divergence or other f-divergences in the future. 

Looking forward, our goal is to refine and apply hallucination analysis methods. We want to explore optimized generative reconstructions for minimal hallucinations. We are also interested to compare various diffusion-based reconstruction methods and evaluate their relative susceptibility to hallucinations.

In conclusion, the analysis of hallucinations in medical imaging is of paramount importance, especially as generative reconstruction models become more prevalent. The proposed Hallucination Index represents one option for an quantitative metric for image quality assessment, offering a tool to navigate the complexities of enhancing image quality while minimizing the introduction of potentially misleading generative artifacts. As we continue to refine this metric and apply it to a broader range of models and applications, we anticipate valuable insights that will contribute to the ongoing improvement of medical image reconstruction.

\newpage

%
%
%
\bibliographystyle{splncs04}
\bibliography{Paper-3513}

\end{document}